# Enhancing creep resistance in refractory high-entropy alloys: role of grain size and local chemical order

Saifuddin Zafar[1], Mashaekh Tausif Ehsan[1+], Sourav Das Suvro[1+], Mohammad Nasim Hasan[1], and Mahmudul Islam[2*]

[1]Department of Mechanical Engineering, Bangladesh University of Engineering and Technology, Dhaka, Bangladesh

[2]Department of Materials Science and Engineering, Massachusetts Institute of Technology, Cambridge, MA, USA

[*]Corresponding authors: ridul@mit.edu

[+]These authors contributed equally to this work.

## Abstract

Refractory high-entropy alloys (RHEAs) are a promising class of materials with potential applications in extreme environments, where the dominant failure mode is thermal creep. The design of these alloys, therefore, requires an understanding of how their microstructure and local chemical distribution affect creep behavior. In this study, we performed high-fidelity atomistic simulations using machine-learning interatomic potentials to explore the creep deformation of MoNbTaW RHEAs under a wide range of stress and temperature conditions. We parametrized grain size and local chemical order (LCO) to investigate the effects of these two important design variables, which are often controllable during the alloy fabrication process. Our findings demonstrate that LCO near grain boundaries can significantly mitigate creep deformation by suppressing grain boundary sliding. This study highlights the importance of utilizing LCO in conjunction with other microstructural properties when designing RHEAs for extreme environment applications.

## Introduction

Materials used in extreme conditions withstand a wide variety of harsh environments that can severely affect their performance and longevity. These environments include elevated temperature and pressure, where materials must maintain structural integrity without succumbing to numerous failure modes. Among the failure modes, thermal creep is the predominant under such conditions. It refers to the gradual time-dependent deformation of materials under constant load at an elevated temperature. The ability of a material to suppress creep failure depends not only on its microstructure (1–4) but also on how its chemistry interacts with the mechanisms that initiate and propagate creep (5–7). Researchers are in constant search of new creep-resistant alloys (4,8–10) to reduce the materials costs associated with high-temperature machinery, such as jet engines and gas turbines. Enhancing creep resistance also reduces the need for frequent part replacements of these machineries. This contributes to environmental sustainability by lowering the emission volume due to materials processing (i.e., casting, annealing, and work-hardening). However, these materials are often constrained by their processing difficulties and limited compositional and microstructural design space.

An emerging candidate class of materials for high temperature applications is high-entropy alloys (HEAs) (11,12). With multiple elements in near-equiatomic proportions, these alloys, often existing in single phase, offer an extremely broad compositional design spectrum, resulting in potentially a limitless variety of materials properties (13–16). The subclass of HEAs that has shown the most promise in terms of creep resistance is known as refractory high-entropy alloys (RHEAs) (17–20). These alloys predominantly consist of refractory elements of the periodic table, such as Nb, V, Zr, Ta, Hf, Mo, and W. These elements have significantly high melting points, allowing their near-equiatomic solid solutions to exhibit superior heat resistance while maintaining exceptional mechanical properties at elevated temperatures (21,22). Several studies have demonstrated that RHEAs can outperform both Ni-based superalloys and traditional refractory alloys under extreme conditions (23,24). However, the challenge lies in designing the microstructure and local chemical distributions of these alloys to achieve the desired mechanical properties at high temperature and pressure. Under extreme conditions, creep failure typically initiates along grain boundaries, which are high-energy regions in the material where cracks can easily propagate (25,26). As creep progresses, the failure mode transitions from intergranular (along the grain boundaries) to transgranular (through the grains themselves) (27,28). The grain size significantly impacts the creep behavior since it determines the total grain boundary area. A smaller grain size results in a larger grain boundary area (29), increasing the likelihood of grain boundary sliding and intergranular diffusion, which often accelerates creep failure. Conversely, a larger grain size reduces the grain boundary area, typically enhancing resistance to grain boundary-related creep mechanisms but possibly promoting transgranular failure (30). In the case of RHEAs, with their pronounced chemical complexity, this behavior may significantly be different than conventional alloys. The nanoscale local chemical distribution, often quantified as local chemical order (LCO), has been shown to notably affect the properties of HEAs (31–37). It is thus reasonable to assume that the local chemical environment near grain boundaries could play a crucial role in determining whether and how creep initiates. This highlights the need for further research on the combined effects of grain microstructure and LCO on thermal creep to tailor these alloys for optimal high-temperature performance.

## Results

### *Creep behavior under different conditions*

We carried out molecular dynamics simulations to study the thermal creep behavior of the paradigmatic refractory high-entropy MoNbTaW alloy using high-fidelity machine learning interatomic potential. The MoNbTaW nanocrystalline structures were generated by randomly placing atoms of different elements on body-centered cubic lattice in equiatomic proportions. These structures, devoid of any chemical order, are referred to as random solid solution (RSS) structures (illustrated in fig. 1a). To simulate the creep behavior of this alloy under extreme environmental conditions, we applied three different temperatures (1500K, 1750K, and 2000K) across three different constant uniaxial stresses (4.5 GPa, 5 GPa, and 5.5 GPa). These conditions correspond to approximately 0.47–0.62 times the melting temperature ($T_m$) and 0.45–0.55 times the yield stress ($\sigma_y$) (See supplementary section 1 and 2). To examine the effect of grain size on the creep behavior of this alloy, we considered three different average grain sizes ($d$): 8.61 nm, 10.21 nm, and 11.72 nm, one of which is illustrated in fig. 1b. At a high temperature, relatively large uniaxial stress induces strain characteristic of thermal creep (fig. 1c).

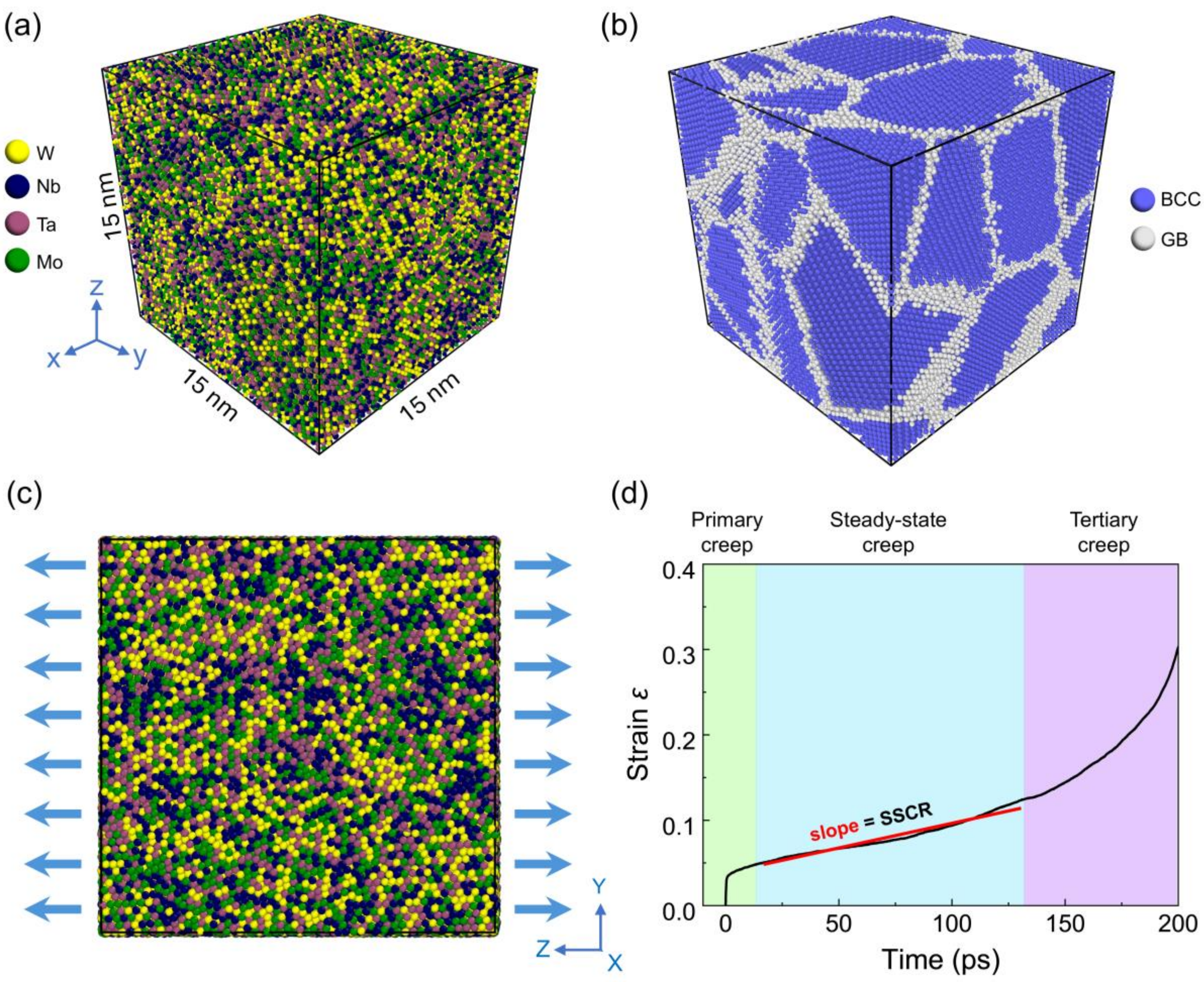

**Figure 1: (a)** Atomic configuration and corresponding **(b)** microstructure of a representative nanocrystalline equiatomic MoNbTaW system. **(c)** Uniaxial stress is imposed along z-direction to initiate creep deformation. **(d)** Time evolution of strain during creep simulation, illustrating three stages of creep.

The time-dependent strain response of a representative system is illustrated in fig. 1d. We observe that the applied stress causes an immediate increase in strain during the initial stage, marking the primary stage of creep, which is characterized by a very high strain rate. Following this stage, strain increases steadily over time, entering the steady-state creep stage, where the creep rate remains relatively constant. The steady-state creep rate (SSCR) is an important indicator of the material's creep resistance: a higher SSCR signifies lower resistance to creep deformation, while a lower SSCR indicates higher resistance to creep. After the steady-state creep stage, the strain rapidly increases, leading to the tertiary creep stage, where the material undergoes rapid deformation and eventually fails.

Figure 2 illustrates the creep profiles of nanocrystalline RSS structures under different simulation conditions. As we can see, the material response under creep loading is highly dependent on the magnitude of the applied stress, temperature, and grain size. The effect of temperature can be observed from any sub-figure column in fig. 2. As expected, higher temperature makes the system more prone to creep failure, as evident by the higher SSCR at these temperatures. For instance, the structure with a grain size of 11.72 nm, subjected to 5.5 GPa stress at 1500K exhibited no tertiary creep stage (fig. 2a) within the simulation timescale. However, when the temperature was increased to 1750K, the same structure under same stress showed a higher SSCR followed by a tertiary creep stage (fig. 2d). This behavior is consistent across all the grain sizes considered in this study. The applied stress level has a similar effect as temperature, which can be observed in any sub-figure row in fig. 2. An increase in stress level results in a higher SSCR, indicating that the material is more susceptible to creeping with increasing stress. This happens because higher stress facilitates dislocation nucleation and grain boundary diffusion, thereby accelerating the creep process (38,39). Additionally, higher stress levels lead to greater strain, which increases the formation of vacancy defects and promotes vacancy diffusion (40).

***Effect of grain size on creep***

We now turn to the effect of grain size on the creep response of RHEA. At lower stress levels and temperatures, the material exhibits only the primary and steady-state creep stages during the simulation timeframe (200 ps). In contrast, at higher stress levels and temperatures, the material progresses through all three stages of creep. Whether or not the material will reach the rapid creep stage is dependent on the grain size. We observe in fig. 2 that across different conditions, the polycrystalline MoNbTaW having larger grain size of 11.72 nm is less likely to undergo the rapid creep stage. On the other hand, the other two grain sizes demonstrate the rapid creep stage even under moderate conditions, such as, at $T = 1500$ K and $\sigma = 5.5$ GPa. This behavior is reminiscent of the so-called inverse Hall-Petch effect (41). The conventional explanation for this behavior is straightforward: structures with smaller grains experience more grain boundary sliding and diffusion-based creep (i.e., Coble creep) due to larger grain boundaries (42), leading to faster deformation under moderate conditions. Larger grains, with fewer grain boundaries, rely on dislocation motion, providing greater resistance to rapid creep, especially at lower stress and temperature(43).

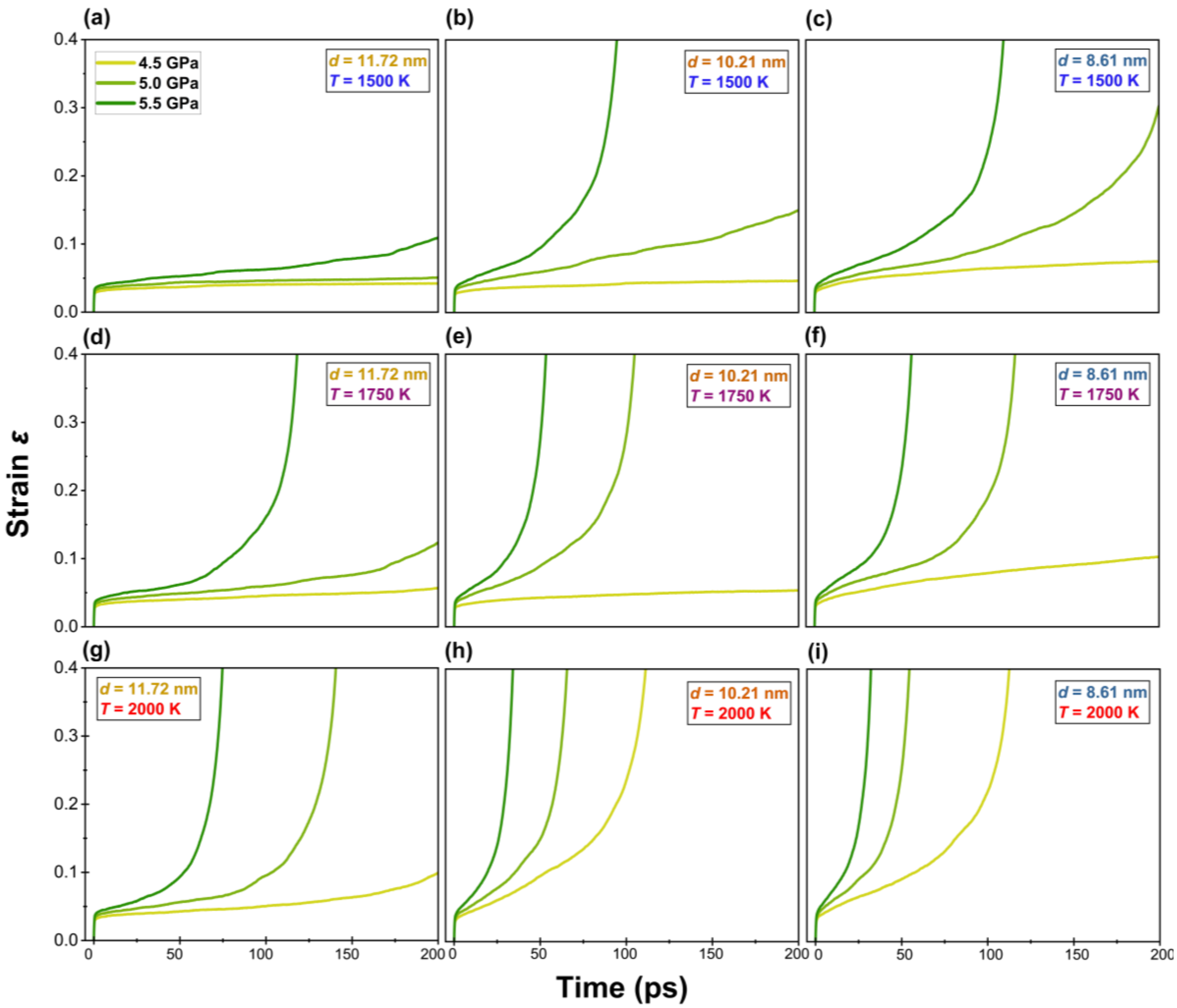

**Figure 2:** Creep profile of nanocrystalline equiatomic MoNbTaW RHEA of different grain sizes (*d*), temperatures (*T*) and applied stresses ($\sigma$).

The underlying mechanism that leads to the creep deformation can be dominated by either grain boundary diffusion, grain boundary sliding, dislocation nucleation or a combination of these mechanisms. To quantitatively analyze this in context of our simulations, we use the classical Bird-Dorn-Mukherjee equation that relates SSCR, $\dot{\epsilon}$, with stress $\sigma$ and grain size $d$ (44,45):

$$\dot{\varepsilon} = A\sigma^n \left(\frac{1}{d}\right)^p e^{-\frac{Q}{kT}} \tag{1}$$

where $A$ is material constant, $\sigma$ is the applied stress, $k$ is the Boltzmann's constant, and $T$ is the absolute temperature. Activation energy, $Q$, is a measure of creep resistance, higher $Q$ will result in greater thermal stability and creep resistance (46,47). The exponent values, $n$ (stress exponent) and $p$ (grain size exponent) provide an estimation of the underlying creep mechanism (48,49). We have extracted the values of $n$ and $p$ from fig. 2, by linearly fitting the relations of $\ln(\dot{\varepsilon})$ with $\ln(\sigma)$ and $\ln\left(\frac{1}{d}\right)$, respectively, as illustrated in fig. 3a-f. Stress exponent, $n$, is greater than 3 for all cases and in some cases, it is greater than 10 which indicates that the material's creep behavior is deviating from the classical power-law creep regime. Such breakdown of power-law creep behavior has been previously reported experimentally at high stress conditions (50). This high value of $n$ indicates a dislocation dominated creep deformation for all the cases (51).

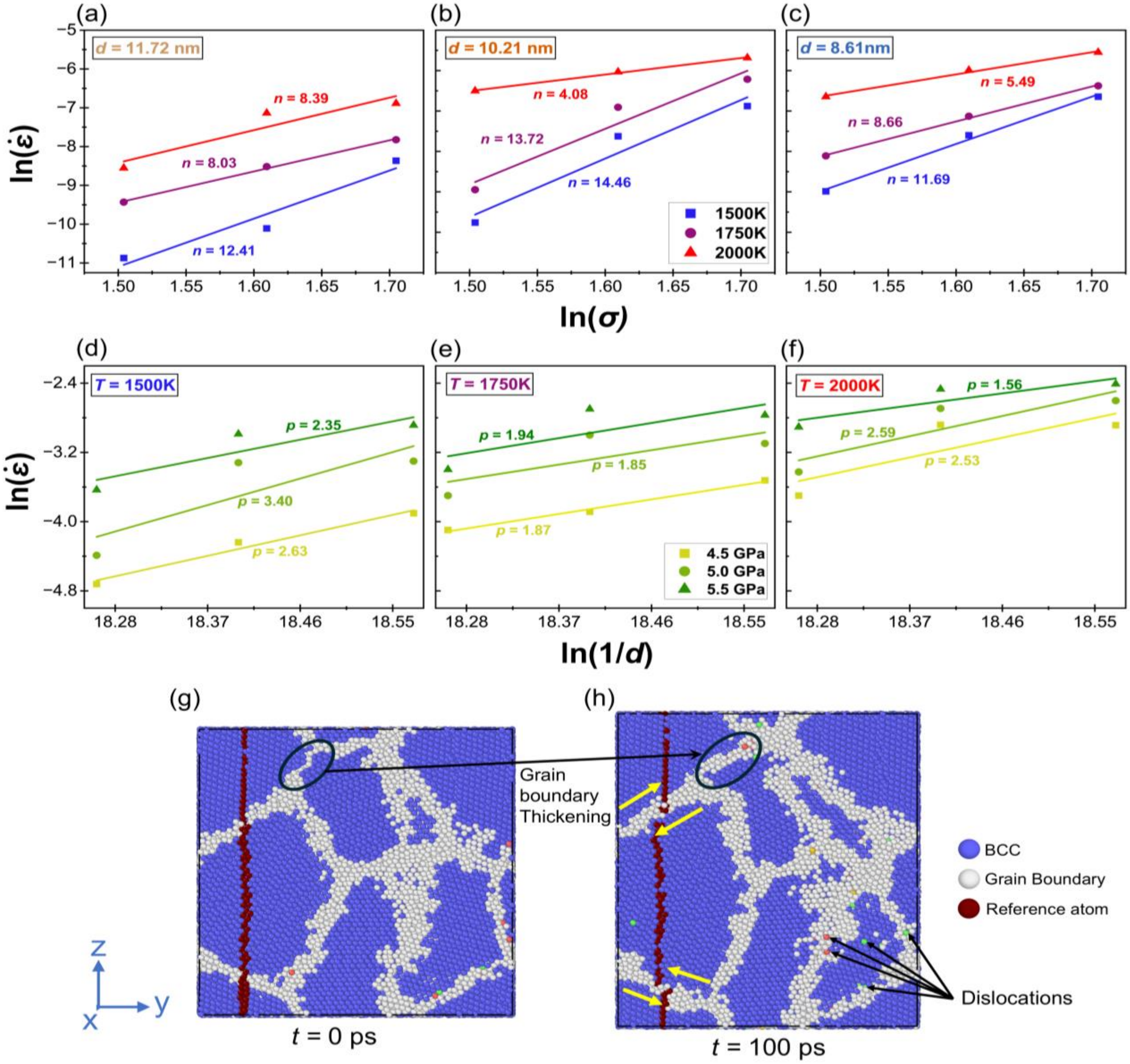

**Figure 3. (a–c)** Effect of uniaxial stress level on the steady-state creep rate, demonstrating that the stress exponent $n$ in eq. 1 is strongly dependent on temperature ($T$) and grain size ($d$). **(d–f)** Effect of grain size on the steady-state creep rate, showing that the grain size exponent $p$ depends on both applied stress ($\sigma$) and $T$. **(g-h)** Grain boundary thickening during creep is illustrated by the increase in the number of non-body-centered cubic (white) atoms in the system. Grain boundary sliding is depicted by the displacement of reference atoms (red) within the grain boundary region.

Additionally, we found that $p$ was close to 2 and within the range of 2 to 3 for most cases. This range indicates that the creep deformation mechanism is primarily governed by a combination of grain boundary sliding and dislocation nucleation (52). Under the high-stress conditions simulated here, dislocations are highly activated, interacting strongly with grain boundaries. These interactions generate localized stress and strain concentrations, which promote atomic rearrangements near the boundaries, leading to increased lattice disorder in these regions (53). This effect is clearly observed in our simulations, where we see both thickening of the grain boundary region and evidence of grain boundary sliding as deformation progresses (fig 3g-h). In this mechanism, the extent of grain boundary area plays a significant role: a larger grain boundary area provides more interface for dislocations to interact, serving both as a source and sink for dislocations (49). Conversely, with larger grain sizes (and therefore less grain boundary area), these interactions are reduced, limiting the extent of grain boundary sliding and dislocation activity. Thus, increasing the grain size reduces grain boundary sliding and suppresses the creep deformation process. This is also demonstrated in fig. 2 where we see that larger grain size results in smaller SSCR.

***Effect of local chemical order on creep***

The MoNbTaW structures simulated and investigated so far in this study are devoid of any local chemical order (LCO). To achieve chemical order in these samples, we performed hybrid Monte Carlo molecular dynamics (MC/MD) simulation that imitates the annealing process at 300K. After the annealing simulation, the structure (8.61 nm grain size) develops LCO. We refer to this structure as LCO. We calculated the Warren-Cowley (WC) parameters, $\alpha$ of both RSS and LCO (same grain size) and observed that the LCO structure has significantly larger WC parameters than its RSS counterparts as shown in fig. 4a. The negative value of $\alpha$ indicates pair preference whereas a positive value means that the pair is less likely to be present in the alloy. $\alpha$ near 0 means no preference of atomic bonding, indicating random solid solution behavior (54). Fig. 4b clearly illustrates the presence of LCO in local chemical distribution in LCO structure compared to the corresponding RSS structure. If we observe fig. 4c, we notice that in LCO, the Nb atoms are segregated near the grain boundary whereas W atoms are accumulated in the grains. Among the four elements present in this alloy, Nb has the lowest grain boundary energy and W has the highest (55). This behavior is consistent across a wide range of tilt and twist boundary conditions. The grain boundary energy thus behaves as a driving force that results in a segregation of Nb near grain boundary. The existence of pronounced LCO near the grain boundary stabilizes the grain boundaries as illustrated by figure 4c. The strain is primarily localized in the grain boundary region, and we observe that the strain at 100 ps near the grain boundary for $T$ = 1500 K and $\sigma$ = 5 GPa, is much higher in RSS whereas the Nb segregation in LCO structure reduces this strain. We can infer from fig. 4c that the grain boundaries in LCO structures are more stable.

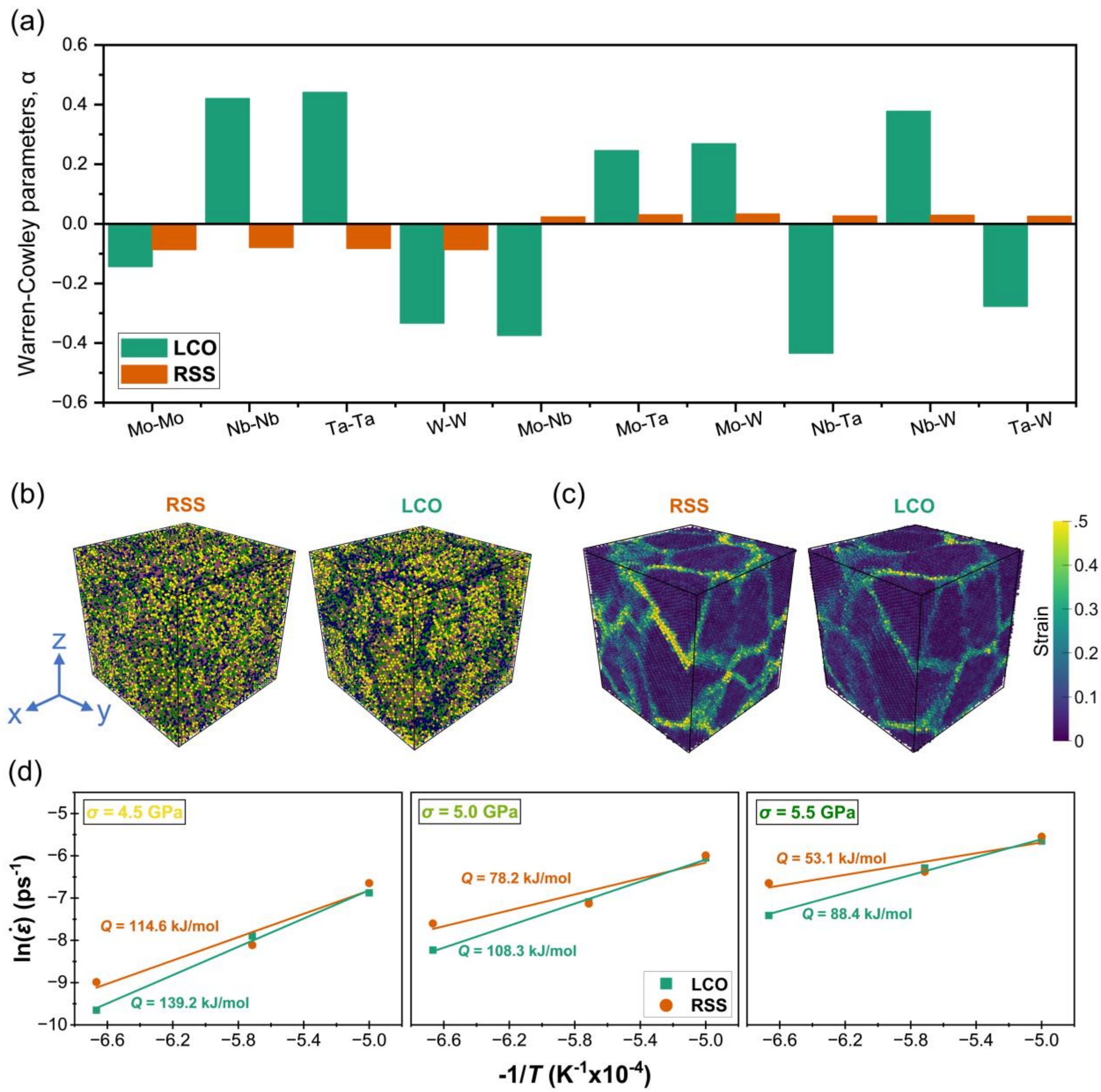

**Figure 4:** **(a)** Warren-Cowley (WC) parameters of random solid solution (RSS) and thermally equilibrated (LCO) MoNbTaW systems. **(b)** Atomic configuration and **(c)** local strain distribution of both RSS and LCO structures. **(d)** Effect of temperature on steady state strain rates under various stress conditions for both RSS and LCO structures.

We studied the effect of LCO on the creep resistance by performing thermal creep simulations on LCO structure under three different stress conditions of 4.5 GPa, 5.0 GPa and 5.5 GPa at three different temperatures, 1500K, 1750K and 2000K. Note that these conditions are the same as the conditions we subjected to the RSS structures. As can be seen from fig. 4d, the SSCR of LCO structure is consistently less than that of RSS under all the conditions demonstrating that LCO does indeed play a role in creep resistance of the alloy. To understand the mechanism by which Nb segregation near the grain boundary (LCO) contributes to creep resistance, we analyzed the activation energy, $Q$, of creep using eq. 1. The activation energy was determined by linearly fitting the relationship between $\ln(\dot{\varepsilon})$ (logarithm of steady state creep rate) and $-1/T$ (inverse of temperature). The slope of this linear fit yields the value of $Q/k$ ($k$ = Boltzmann's constant), which directly governs the rate of creep: a higher $Q$ value indicates a larger energy barrier for creep, resulting in greater resistance to deformation (56).

Figure 4d compares the activation energy, $Q$, for the RSS and LCO structures across various tensile stress. For all stress levels, the LCO structure exhibits consistently higher $Q$ values compared to the RSS structure, highlighting the enhanced creep resistance provided by LCO. For instance, under an applied stress of 4.5 GPa, $Q$ for the LCO structure is 139.6 kJ/mol, while for the RSS structure, it is only 114.6 kJ/mol. This increase in $Q$ is attributed to the stabilization of grain boundaries in the LCO structure, both chemically (through Nb segregation) and structurally (through reduced grain boundary strain). However, as the temperature increases, the difference in $Q$ between the LCO and RSS structures diminishes. This trend suggests that the effectiveness of LCO in enhancing creep resistance is temperature dependent. At lower temperatures, creep is primarily governed by grain boundary sliding, a diffusion-controlled process. The stabilization of grain boundaries by LCO effectively impedes grain boundary sliding, resulting in enhanced resistance to creep in the LCO structure. This stabilization arises from Nb segregation at grain boundaries, which lowers grain boundary energy and hinders the sliding process. In contrast, at higher temperatures, creep is dominated by dislocation-mediated mechanisms such as climb and glide, where grain boundary stabilization through LCO is less effective. As a result, the resistance provided by LCO diminishes at elevated temperatures.

## Discussion

Our investigation of creep deformation in refractory high-entropy alloys provides new insights into the role of their chemical complexity in controlling creep behavior. Specifically, we observe that in RHEAs such as MoNbTaW, the chemical complexity induces pronounced local chemical order near the grain boundaries. This ordering is characterized by the preferential accumulation of specific chemical species at the grain boundaries, which is driven primarily by the grain energy associated with the constituent elements. The grain boundary ordering observed in our simulations leads to an increase in the activation energy for creep deformation (fig. 4d). This stabilization effect is particularly pronounced in cases where grain boundary sliding is the dominant creep mechanism. Grain boundary sliding, facilitated by the relative motion of grains along their boundaries, is typically more significant in finer-grained systems. However, our results indicate that the enhanced local chemical order near the grain boundaries acts to suppress sliding by "locking" the boundaries and restricting their motion.

The suppression of grain boundary sliding through local chemical ordering suggests a novel design strategy for improving the high-temperature creep resistance of RHEAs. By tailoring the alloy composition to promote chemical ordering at grain boundaries, it may be possible to achieve superior creep resistance even in materials with relatively small grain sizes. The material can then demonstrate both superior ductility and creep resistance, which are highly desirable properties for extreme environment applications. This approach could involve selecting elements with specific grain energy characteristics to maximize the beneficial effects of chemical ordering and grain boundary stabilization. Our study thus suggests that both LCO and microstructure should be considered as co-design parameters for developing RHEAs intended for extreme applications.

## Methods

### *Interatomic potentials*

Machine learning spectral neighbor analysis potential (ML-SNAP) was used to describe the interatomic interactions in all the atomistic simulations performed in this study. This potential has been specifically trained to replicate the chemical short-range order in MoNbTaW alloy (57). Furthermore, the potential has been validated to accurately predict important materials properties such as cohesive energy, mechanical characteristics, and relative phase stability, critical for our present simulations. Large-scale Atomic/Molecular Massively Parallel Simulator (LAMMPS) (58) was used to carry out all the atomistic simulations.

### *Warren-Cowley parameters*

We quantified local chemical order in the thermally equilibrated structures by adopting the widely used Warren-Cowley (WC) parameters (54), defined as:

$$\alpha_{ij} = 1 - \frac{p_{ij}}{c_j} \quad (2)$$

where *i* and *j* refer to the elements, $c_j$ is the nominal concentration of *j* type atom, and $p_{ij}$ represents the fraction of type *j* in the nearest neighboring shell of *i* type atom. $\alpha \approx 0$ indicates a completely random distribution of *i* and *j* atoms. A negative value of $\alpha$ signifies clustering of *j* type atoms around *i* type atoms, whereas a positive value indicates a tendency for *j* type atoms to avoid *i* type atoms in their nearest neighbor shell.

### *Hybrid MC/MD simulation*

Nano-polycrystalline MoNbTaW structures with equi-atomic chemical composition, considered in this study were generated by the Voronoi algorithm (59) using ATOMSK (60). Each of these structures had dimensions of 15 nm × 15 nm × 15 nm, containing approximately 172,000 atoms as illustrated in fig. 1a. Here, we considered three different average grain sizes (8.61 nm, 10.21 nm, and 11.72 nm) with periodic boundary conditions in all three directions. These structures were initialized by randomly distributing all four elements in equi-atomic proportions. We have referred to these structures as random solid solution (RSS) structures. We thermally equilibrated the RSS structure with the average grain size of 8.61 nm, by performing hybrid Monte Carlo molecular dynamics (MC/MD) simulations. In this simulation, each MC swap involved swapping of atoms of different element types based on the Metropolis algorithm (61). The acceptance of these atom swaps was determined by the simulation temperature, which was kept 300 K. After every 10 MC swaps, the structure was relaxed at 300K and zero pressure using the NPT ensemble with damping coefficient of 1 ps. The timestep for the MD relaxation was 1 fs. A total of 100,000 MC/MD steps were carried out which includes 1 million MC swaps and 100 ps of MD relaxation. At the end of the simulation, the annealed MoNbTaW structure displayed pronounced local chemical preference and much lower potential energy, as opposed to its RSS counterpart, which had near-zero WC parameters. We have referred to this structure as local chemical ordered (LCO) structure.

***Creep Simulation***

We performed molecular dynamics simulations of thermal creep under uniaxial tension. The simulation cell (of MoNbTaW alloy) was first minimized energetically using conjugate gradient (cg) method and then equilibrated for 50 ps at the creep temperature and zero pressure under NPT ensemble with damping coefficient of 1 ps. After equilibration, a constant tensile stress was applied along the z-direction with zero pressure applied along x- and y-directions under NPT ensemble for 200 ps. Three different stress levels and three different temperatures were chosen, which totals to nine different loading conditions. The selected creep temperatures (1500K, 1750K and 2000K) were kept well below the melting temperature (3210K). See supplementary section 1 for the detail on the melting point calculation. The considered stress levels (4.5 GPa, 5.0 GPa and 5.5 GPa) fall below the yield stress of the material at the highest creep temperature (2000K). See Supplementary section 2 for detailed calculation of yield stress from the stress-strain curve. The timestep of the simulation was 1 fs.

The evaluation of microstructural configuration and dislocation analysis of the structure during the creep simulation were evaluated using the common neighbor analysis (CNA) (62) and dislocation analysis (DXA) (63), respectively, in Ovito (64).

## Data and code availability

The data that support the findings of this study are available from the corresponding authors upon reasonable request.

## Acknowledgement

The authors acknowledge the MIT SuperCloud and Lincoln Laboratory Supercomputing Center and Institute of Information and Communication Technology (IICT) of Bangladesh University of Engineering and Technology (BUET) for providing high-performance computational resources that have contributed to the research results reported within this paper. The authors also acknowledge the Department of Mechanical Engineering, BUET, for their support for this project. M.I. is supported by MathWorks Engineering Fellowship Fund.

## Author contributions

S.Z., M.T.E., S.D.S., M.I. and M.N.H. conceived the project. S.Z. conducted the hybrid MC/MD simulations. S.Z. performed the creep simulations. S.Z and M.I. wrote the original manuscript. All authors contributed to the interpretation of the results, prepared, reviewed, and edited the manuscript. Project supervision was done by M.I. and M.N.H. Project administration was performed by M.N.H.

## Competing interests

The authors declare no competing interests.

# Supplementary Information

## Enhancing creep resistance in refractory high-entropy alloys: role of grain size and local chemical order

Saifuddin Zafar[1], Mashaekh Tausif Ehsan[1+], Sourav Das Suvro[1+], Mohammad Nasim Hasan[1], and Mahmudul Islam[2*]

[1]Department of Mechanical Engineering, Bangladesh University of Engineering and Technology, Dhaka, Bangladesh

[2]Department of Materials Science and Engineering, Massachusetts Institute of Technology, Cambridge, MA, USA

[*]Corresponding authors: ridul@mit.edu

[+]These authors contributed equally to this work.

## Supplementary Section 1: Melting point calculation

We determined the melting point of the equiatomic MoNbTaW alloy using the solid-liquid coexistence method (1–4). As shown in supplementary fig. 1a, we constructed a simulation cell comprising two equal halves: one consisting of 12,116 solid BCC atoms and the other of 12,116 liquid atoms. Each subsystem was independently minimized using the conjugate gradient algorithm to ensure no overlap of atoms. Following this, the entire system was equilibrated for 50 ps at various temperatures using the NPT ensemble. The timestep for the simulation was set to 1 fs, with a damping coefficient of 1 ps. The results for different temperature cases show varying fractions of solid and liquid, as illustrated in supplementary fig. 1. Supplementary table 1 indicates that at 3220 K, the percentages of solid and liquid closely match the initial configuration, suggesting this temperature is approximately the melting point of the alloy.

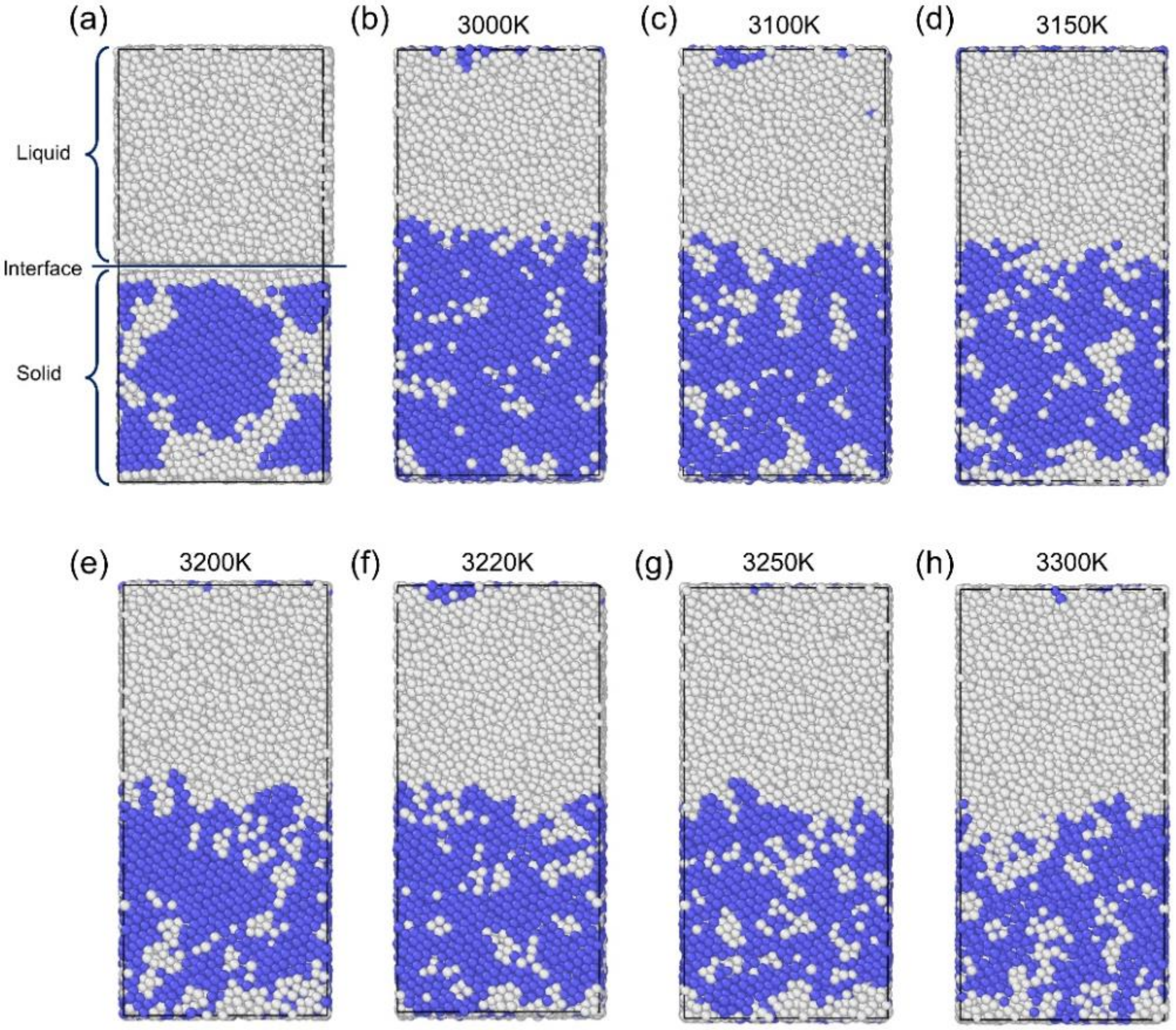

**Supplementary Figure 1:** Atomic configuration of solid and liquid phase after equilibration at different temperatures. Atoms are colored using common neighbor analysis. Blue atoms represent perfect BCC atoms.

**Table 1:** Portion of solid phase and liquid phase after thermal equilibration at different temperatures. Atoms in BCC phase are considered solid phase and amorphous atoms are considered liquid phase.

| **Temperature (K)** | Initial | 3000 | 3100 | 3150 | 3200 | 3220 | 3250 | 3300 |
|---|---|---|---|---|---|---|---|---|
| **BCC (Solid) (%)** | 36.6 | 48.9 | 45.2 | 41.4 | 41.1 | 37.5 | 38.2 | 35.1 |
| **Others (Liquid) (%)** | 66.7 | 51.1 | 54.8 | 58.6 | 58.9 | 62.5 | 61.8 | 64.9 |

## Supplementary Section 2: Yield Stress calculation

We conducted uniaxial tensile simulations on a nano-crystalline MoNbTaW structure with dimensions of 15 nm × 15 nm × 15 nm and average grain sizes of 8.61 nm, 10.21 nm, and 11.72 nm. Periodic boundary conditions were applied along the x, y, and z directions, and a strain rate of $10^9$ $s^{-1}$ was imposed. The resulting stress-strain curves are presented in supplementary fig. 2. The yield strength was determined by drawing a line parallel to the elastic region of the stress-strain curve at an offset of 0.2% strain. The intersection of this offset line with the stress-strain curve indicates the yield strength. Across all grain sizes, the yield stress was consistently found to be approximately 10.5 GPa. These calculations were performed at a temperature of 2000 K, the maximum temperature at which creep simulations were conducted.

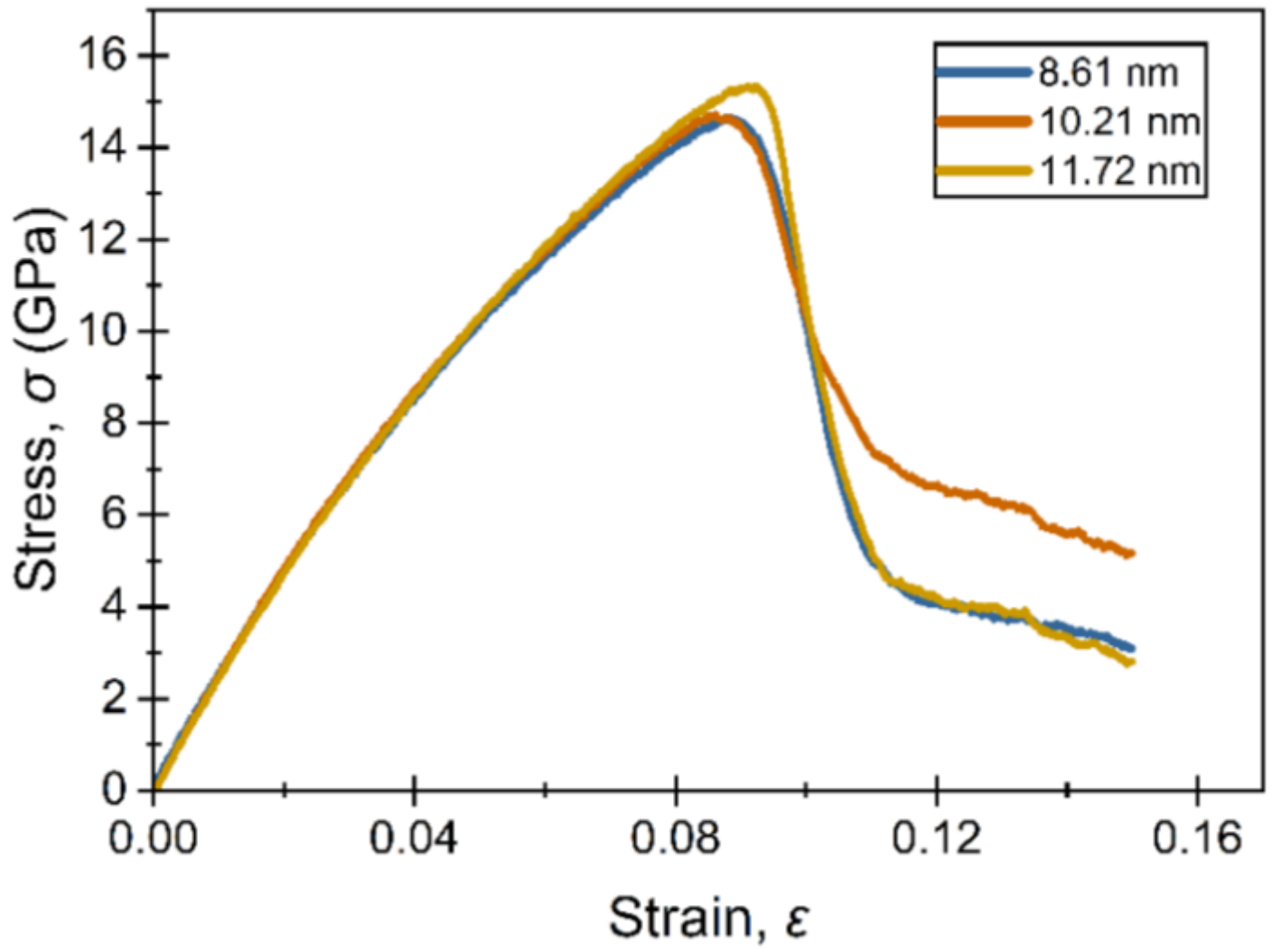

**Supplementary Figure 2:** Stress-strain curve of MoNbTaW nanocrystalline structure.